\documentclass[10pt]{article}
\usepackage{amsfonts}
\usepackage{amssymb}
\usepackage{amsthm}
\usepackage{amsmath} 
\usepackage{graphicx}
\usepackage{color}
\date{}
\newcommand{\sss}{\setcounter{equation}{0}}
\newtheorem{theorem}{THEOREM}[section]

\newtheorem{lemma}[theorem]{LEMMA}

\newtheorem{remark}[theorem]{REMARK}

\newtheorem{prop}[theorem]{PROPOSITION}

\newtheorem{example}[theorem]{EXAMPLE}

\def\beq{\begin{equation}}
\def\ene{\end{equation}}

\newcommand{\bull}{\hfill $\Box$}

\def\stretch {\noalign{\medskip}}

\def \bm {\bmatrix}
\def\endbm {\endbmatrix}

\begin{document}
\baselineskip=20 pt
\parskip 6 pt

\title{Trace Identities  for the  matrix Schr\"odinger operator on the half line with general boundary conditions
\thanks{ PACS classification (2010) 02.30.Zz; 03.65.-w; 03.65.Ge; 03.65.Nk. Mathematics Subject Classification (2010): 34L25; 34L40; 81U05; 81Uxx. Research partially supported by project  PAPIIT-DGAPA UNAM  IN102215}}
\author{ Ricardo Weder\thanks {Fellow, Sistema Nacional de Investigadores.}\thanks{ Electronic mail: weder@unam.mx. Home page: http://www.iimas.unam.mx/rweder/rweder.html} \\
Departamento de F\'{\i}sica Matem\'atica.\\
 Instituto de Investigaciones en Matem\'aticas Aplicadas y en
 Sistemas. \\
 Universidad Nacional Aut\'onoma de M\'exico.\\
  Apartado Postal 20-126,
M\'exico DF 01000, M\'exico.}

\maketitle

\vspace{.5cm}
 \centerline{{\bf Abstract}}
  We prove Buslaev-Faddeev trace identities  for the matrix Schr\"odinger operator  on the half line, with  general boundary conditions at the origin, and with  selfadjoint matrix  potentials.

\bigskip

\section{Introduction}\sss
In this paper we study  the matrix Schr\"odinger operator on the half line
\begin{equation}
H_{A,B}\psi:= -\psi''+V(x)\,\psi,\qquad x\in(0,\infty),\label{1.1}
\end{equation}
where the prime denotes the derivative with respect to
the spatial coordinate $x$. Furthermore, the wavefunction $\psi(x)$ 
will  either be an $n\times n$ matrix-valued function
or it will be a column vector valued function with $n$
components. As it is shown in  \cite{11}-\cite{wes}  the  general selfadjoint boundary conditions at $x=0$ for the matrix Schr\"odinger operator \eqref{1.1} can be formulated in several equivalent way.  For our purposes, it is more convenient to  use the formulation given in in \cite{19}-\cite{wes} where we state them  in terms of constant $n\times n$ matrices $A$ and $B$ as follows, 
\begin{equation}-B^\dagger\psi(0)+A^\dagger\psi'(0)=0,\label{1.2}
\end{equation}
\begin{equation}-B^\dagger A+A^\dagger B=0,\label{1.3}\end{equation}
\begin{equation}A^\dagger A+B^\dagger B>0.\label{1.4}\end{equation}
Note that  $A^\dagger B$ is selfadjoint and
the selfadjoint matrix $(A^\dagger
A+B^\dagger B)$ is positive.

It is clear that the  matrices $A, B$ are not uniquely defined. It is always possible to multiply them on the right by an invertible matrix $T$ without affecting  \eqref{1.2}, \eqref{1.3} and \eqref{1.4}.  Moreover,
\beq\label{1.5}
H_{A\,T, B\,T}= H_{A,B}.
\ene

We assume that the potential matrix $V(x)$ is a $n\times n$ selfadjoint matrix-valued function that is in the Faddeev class, namely, $V(x)$ and its first moment  are  integrable in $(0, \infty)$. That is to say,  each entry of the matrix $V$ is Lebesgue measurable on
$(0, \infty)$ and
\begin{equation}
\int_0^\infty\, ( 1+x)\, 
dx\,||V(x)||<+\infty.\label{1.6}
\end{equation} 
By ,  $||V(x)||$ we  denote   the norm of  $V(x)$ as an operator on $\mathbf C^n$.
Clearly, $V(x)$  satisfies  \eqref{1.6} if and only  this equation holds for each of its entries. Moreover, we assume that  the potential is infinitely differentiable on $(0,\infty)$  and that,
\beq \label{1.7}
\left\|  \frac{d^j}{d x^j}\, V(x)  \right\| \leq C_j \, (1+|x|)^{-\rho-j}\, \textrm{for some}\, \rho \in (1,2], \, \textrm{and all}\, j=0,1,2,\cdots.
\ene

Furthermore, we always assume that the matrix potential is selfadjoint ( the dagger means matrix adjoint) 

\begin{equation}
V(x)=V(x)^\dagger,\qquad x\in (0, \infty).\label{1.8}
\end{equation}
The main result of our paper is Theorem 3.6 where we prove  Buslaev-Faddeev  trace identities for the matrix Schr\"odinger operator \eqref{1.1} with the most general  boundary conditions (\ref{1.2}-\ref{1.4}). These trace identities were first proven in the scalar case by Buslaev and Faddeev in \cite{bf} (see also  \cite{f} ). For a textbook presentation of these results see   Section 6 of Chapter 4 of \cite{ya2}. In these remarkable identities the sum of the  absolute value of the eigenvalues, to an even or to an odd power, is expressed in terms   of the coefficients of the asymptotic expansion for high energy of the logarithm of the Jost function and of the integral over the absolutely continuous spectrum $(0,\infty)$  of the phase of the Jost function in the even case and of the logarithm of the absolute value of the Jost function in the odd case.   The coefficients on the asymptotic expansion of the logarithm of the Jost function depend on the (scalar) potential. These identities give  formulae for sums of the absolute value of the eigenvalues to an even or odd power (traces)  in terms of  properties of the absolutely continuos spectrum encoded in the Jost function. Actually, they link the point  spectrum with the absolutely continuous spectrum, or in other words, bound state information with scattering information.  In \cite{21} we proved  Levinson's theorem for the matrix Schr\"odinger operator with general boundary conditions, that is a trace identity of order zero.    

We prove our trace identities adapting to the  matrix Schr\"odinger operator the classical proof in the scalar case that is given, for example, in Section 6 of Chapter  4 of \cite{ya2}. In our case the Jost function is replaced by the determinant of the Jost matrix. The new technical results that made possible to adapt the classical proof to the matrix case are the precise study of the low-energy behavior of the determinant of the Jost matrix that we obtained in Corollary 6.2 of \cite{21} assuming that \eqref{1.6} holds, the detailed analysis of the high-energy behavior of the determinant of the Jost matrix that is given in Proposition 7.5 of \cite{21} assuming only that the matrix potential is integrable, i.e., that
\beq \label{1.9}
\int_0^\infty\, 
dx\,||V(x)||<+\infty.
\ene    
Furthermore, we use to prove Theorem 3.6 the results of Theorems 8.1 and 8.5 of \cite{21} where, in particular, we prove that the eigenvalues of $H_{A,B}$ coincide with the zeros of the determinant of the Jost matrix and that  the multiplicity of each eigenvalue is equal to the order of the corresponding  zero of the determinant of the Jost matrix. Furthermore, we also use in the proof of Theorem 3.6 the asymptotic expansion for high energy for the logarithm of the determinant of the Jost matrix, that we prove in Section 3 of this paper,  assuming that   $V(x) \in C^\infty((0,\infty))$ and that \eqref{1.7} holds.

 Currently there is a great  deal of interest in the spectral and scattering theory of matrix Schr\"odinger operators on the half-line with general boundary conditions. For a review of the literature see   \cite{19}-\cite{wes}. They are important, for example,  in the quantum mechanical scattering of particles with internal structure and in quantum graphs. 
 The matrix Schr\"odinger operator  on the half line \eqref{1.1} is equivalent  to   a star graph, i.e. to a quantum graph with only one vertex and a finite number of edges of infinite length. Physically it corresponds  to  $n$  very thin quantum wires connected at the vertex. The boundary  conditions \eqref{1.2}, \eqref{1.3} and \eqref{1.4}
impose  relations, at the vertex,    between the values of the wave functions, and of its derivatives, at different edges. See for example,     \cite{11}- \cite{10} ans \cite{5}- \cite{18}

The paper is organized as follows. In Section~2 we state results from \cite{21}, \cite{20}, that we need,  about the Jost solution, the regular solution, the Jost matrix, and in transformations of the matrices $A,B$ that give the boundary conditions. In Section~3 we prove our trace identities and in Section~4 we give examples that illustrate them. 

Along the paper we designate  by  ${\bf C}^+$ the upper-half complex plane,  by ${\bf R}$  the real axis, and we let ${\overline{{\bf C}^+}}:={\bf C}^+\cup{\bf R}$.  For any $k \in  \overline{{\bf C}^+}$ we denote by $k^*$ its complex conjugate. As we already mentioned, for any matrix $D$ we designate by $D^\dagger$ its adjoint. We denote by $C$ a positive constant that is not required  take the same value when it appears in different places.

\section{Preliminary results}\sss

In this section we introduce certain results that we need. See \cite{21} and \cite{20}. We always assume that  the selfadjoint matrix potential $V$ satisfies at least \eqref{1.9}. 
We will use  $n\times n$ matrix solutions to the equation  
 \begin{equation}
 -\psi''+V(x)\,\psi = k^2\, \psi,\qquad x\in(0,\infty), k \in \overline{{\bf C}^+}.\label{3.1}
\end{equation}
Let  $F,G$ be any pair of $n\times n$ matrix valued functions  defined for  $ x \in (0,\infty)$. The Wronskian, [F;G]  is defined as follows,

$$
[F;G]:=FG'-F'G.
$$
 Remark that for any two $n\times n$ solutions $\phi(k,x)$ and
$\psi(k,x)$ to \eqref{3.1}, each of the Wronskians $[\phi(k^*,x)^\dagger;\psi(k,x)]$ and $[\phi(-k^*,x)^\dagger;\psi(k,x)]$ is independent of $x.$

By $f(k,x)$ we denote the Jost solution to \eqref{3.1}  that is the $n\times n$ matrix solution that satisfies the following asymptotics   for 
$k\in{\overline{{\bf C}^+}}\setminus\{0\},$ 

\begin{equation}f(k,x)=e^{ikx}[I_n+o(1/x)],\quad
f'(k,x)=ik\,e^{ikx}[I_n+o(1/x)],\qquad x\to+\infty,\label{3.2}
\end{equation}
with  $I_n$  the $n\times n$ identity matrix. For each fixed $x,$ (see  \cite{21,20})
$f(k,x)$ and $f'(k,x)$ are analytic for $k\in{\bf C}^+$
and continuous for $k\in{\overline{{\bf C}^+}}$. Clearly, this asymptotics implies    that for each fixed $k \in{\bf C}^+$, each of the $n$ columns of
$f(k,x)$  decays exponentially to zero as $x\to+\infty$. 

Another important solution to \eqref{3.1} is the regular solution, $\varphi_{A,B}(k,x),$  that is the $n\times n$ matrix solution 
defined by the   the initial conditions
\begin{equation}
\varphi_{A,B}(k,0)=A,\quad \varphi_{A,B}'(k,0)=B,\label{3.5}
\end{equation}
with  $A$ and $B$  the matrices that define the boundary conditions  in \eqref{1.2}, \eqref{1.3}, \eqref{1.4}.  $\varphi_{A,B}(k,x)$ is entire in $k$
in the complex plane ${\bf C},$ for each fixed $x\in (0,\infty).$ 


Let us define the  Jost matrix $J(k)$ in the following way,
\begin{equation}
J_{A,B}(k):=[f(-k^*,x)^\dagger;\varphi_{A,B}(k,x)],\qquad k\in{\overline{{\bf C}^+}}.\label{3.7}
\end{equation}
 It follows from \eqref{3.7} and evaluating the Wronskian at $x=0$,  that,

\begin{equation}
J_{A,B}(k)=f(-k^*,0)^\dagger B-f'(-k^*,0)^\dagger A,  \qquad k\in{\overline{{\bf C}^+}}.      \label{3.8}
\end{equation}
  $J(k)$ is well defined for ${\overline{{\bf C}^+}}$ since $f(-k^*,0)^\dagger$
and $f'(-k^*,0)^\dagger$ are analytic in $k\in{{\bf C}^+}$ and
continuous in $k\in\overline{{\bf C}^+}.$ It is known \cite{21} that  $J(k)$ is invertible for $ k \in \mathbf{R}\setminus 0$.
 
It is  proven in Proposition 4.1 of \cite{21} (with $M, M^\dagger$ there , replaced, respectively, by $ M^\dagger, M$ ) that under the unitary transformation $V\mapsto M V M^\dagger,$  with a unitary matrix $M,$ and the combination of three consecutive
transformations $(A,B)\mapsto (M AT_1 M^\dagger T_2,M B T_1 M^\dagger T_2),$
first by a right multiplication by an invertible matrix
$T_1,$ then by the unitary transformation with $M,$
followed by a right multiplication by an invertible matrix
$T_2,$ we have  that 
\begin{align}
f_{ M V M^\dagger}(k,x)&= M f(k,x) M^\dagger,  \qquad k \in \overline{{\bf C}^+}, \label{3.9} \\
\varphi_{M\, V M^\dagger, M AT_1 M^\dagger T_2, M BT_1 M^\dagger T_2 }(k,x)  &=M \varphi_{V, A,B}(k,x)
T_1M^\dagger T_2, \qquad k \in {\bf C}, \label{3.10}\\
J_{M\, V M^\dagger, M AT_1 M^\dagger T_2, M BT_1 M^\dagger T_2}(k,x)&= M J_{V,A,B}(k,x)\, T_1 M^\dagger T_2,   \qquad k \in    \overline{{\bf C}^+}.   \label{3.11}
\end{align}
Furthermore, we have that,
\beq \label{3.12}
H_{{ M V M^\dagger, M AT_1 M^\dagger T_2, M B T_1 M^\dagger T_2}} = M \, H_{V,A,B}\, M^\dagger.
  \ene

For clarity, in \eqref{3.9}-\eqref{3.12} we explicitly  state the dependence in $V$ of the Jost solution, the regular solution, the Jost matrix and  the  matrix Schr\"odinger operator.

 The transformation $V\mapsto V$ and $(A,B)\mapsto (AT,BT)$ with
an invertible matrix $T$ is just a change in the 
parametrization of the boundary conditions \eqref{1.2}, \eqref{1.3} and \eqref{1.4}.
On the  contrary, the unitary
transformation $V\mapsto M\, V M^\dagger$ and $(A,B)\mapsto (M A M^\dagger, M B M^\dagger)$
with a unitary matrix $M$ is a change of representation in quantum mechanical sense.

 Taking into account the  general selfadjoint boundary condition  for the  scalar Schr\"odinger operator, that is to say when  $n=1$, we consider  the case where the matrices $A,B$ are diagonal. This special pair   of diagonal matrices is denoted by  $\tilde A$ and $\tilde B,$ with,

\begin{equation}\tilde {A}:=-\mathrm{diag}\{\sin\theta_1,\dots,\sin\theta_n\},
\quad \tilde B:=\mathrm{diag}\{\cos\theta_1,\dots,\cos\theta_n\}.\label{3.12b}
\end{equation}
In this case the boundary conditions \eqref{1.2} are,
\beq\label{3.13}
\cos\theta_j\, \psi_j(0)+ \sin\theta_j\, \psi_j'(0)=0, \qquad j=1,2,\cdots,n.
\ene

Here,  the real parameters $\theta_j$ take values in the interval $(0,\pi].$
The case $\theta_j=\pi$ corresponds to the Dirichlet boundary condition,   the case where $ \theta_j \neq  \pi/2, \pi$ corresponds to mixed boundary conditions and the case $\theta_j=\pi/2$ corresponds to the Neumann boundary condition. We assume  that there $n_{\mathrm D}$ values with $\theta_j=\pi,$ that there 
are $n_{\mathrm M}$ values    $ \theta_j \neq  \pi/2, \pi$ and, in consequence, that  there are $n_{\mathrm N}$ remaining values,
with $n_{\mathrm N}:=n-n_{\mathrm M}-n_{\mathrm D},$ such that the corresponding
$\theta_j$-values are  $\theta_j=\pi/2$. Particular cases where  any of $n_{\mathrm D},$ $n_{\mathrm M},$ and $n_{\mathrm N}$ are  zero or $n$ are allowed. 
Clearly,  $\tilde{A}, \tilde{B}$ satisfy  \eqref{1.3}, \eqref{1.4} with $\tilde{A}, \tilde{B}$ instead of $A,B.$

In Proposition 4.3 of \cite{21} it is proven that for any pair of matrices $(A,B)$ that satisfy \eqref{1.2}-\eqref{1.4} there is a pair of diagonal matrices $(\tilde{A}, \tilde{B})$ as in \eqref{3.12b}, a unitary matrix $M$ and a two invertible matrices $T_1,T_2$ such that,
\beq \label{3.14}
A=    M\, \tilde{A}\, T_2 \, M^\dagger\, T_1, \quad  B=    M\, \tilde{B}\, T_2 \, M^\dagger\, T_1.
\ene
Note that $ T_1,T_2$ in \eqref{3.14} correspond, respectively to  $  T_1^{-1},T_2^{-1}$ in Proposition 4.3 of \cite{21}. 

 The following results are proven in Proposition 6.1 of \cite{21}. Consider the selfadjoint matrix
Schr\"odinger operator \eqref{1.1} with the selfadjoint boundary conditions
(\ref{1.2}, \ref{1.3}, \ref{1.4})  and with the selfadjoint potential  $V$  in the Faddeev class \eqref{1.6},
 Then,  the nonzero column vector
$u\in \bold C^n$ is an eigenvector of the zero-energy Jost matrix $J(0)$
with the zero eigenvalue, i.e.
$u\in \text{Ker}\, [J(0)]$ if and only if
$\varphi_{A,B}(0,x)\,u$ is bounded for $x\in (0,\infty).$ Let us denote by  $\mu$  the geometric multiplicity of the zero
eigenvalue of $J(0).$  Then,
we can form exactly $\mu$ columns by using linear
combinations of $n$ columns of
the zero-energy regular solution $\varphi_{A,B}(0,x)$ in such a way that those
$\mu$ columns form linearly independent solutions to  the zero-energy Schr\"odinger equation,

\beq\label{3.16}
-\varphi''(x)+ V(x)\, \varphi(x)=0,
\ene
that satisfy  the boundary conditions  (\ref{1.2}, \ref{1.3}, \ref{1.4}) and that 
remain bounded as $x\to+\infty.$ Furthermore, each of such $\mu$ column-vector
solutions to \eqref{3.16}  can also be expressed
as a linear combinations of columns of $f(0,x).$ In the physics literature the bounded solutions to \eqref{3.16} that satisfy the boundary conditions are called half-bound states or zero energy resonances. Hence, $\mu$ is the number of linearly independent half-bound states or of zero energy resonances. 

Furthermore, in Corollary 6.2 of \cite{21} it is proven that if the selfadjoint potential is in the Fadeev class \eqref{1.6},  the determinant of
the Jost matrix
$J(k)$ defined in \eqref{3.7} has the small-$k$ behavior
\beq \label{3.17}
\det J(k)=c_1k^\mu[1+o(1)],\qquad k\to 0 \text{ in }\overline{{\bold C^+}},
\ene
 where $c_1$ is a nonzero constant that it is explicitly given in \cite{21}. 

Furthermore, it is proven in Proposition 7.5 of  \cite{21} that if the selfadjoint potential is integrable, i.e. if it satisfies \eqref{1.9}, then,
\beq \label{3.18}
\det J(k)=c_2 k^{n_M+n_N}[1+O(1/k)],\qquad k\to \infty\text{ in }\,\overline{{\bold C^+}},
\ene
 where $c_2$ is a nonzero constant that it is explicitly given in \cite{21},
and $n_M$ and $n_N$ are the nonnegative integers defined
after \eqref{3.13}, i.e. they are, respectively, the number of mixed and of Neumann boundary conditions in the representation where the matrices that define the boundary conditions are given by the diagonal matrices $\tilde{A}, \tilde{B}.$

The following results concerning eigenvalues of $H_{A,B}$  are proven in \cite{21}. Since the matrix Schr\"odinger operator, $H_{A,B}$ is selfadjoint its eigenvalues,  $k^2$,  have to be real. It turns out that $H_{A,B}$ has no eigenvalues for $k^2 \geq 0$ (see the beginning of Section VIII of \cite{21}). Then, the eigenvalues appear at $k^2$, for $ k= i \kappa$, for some $ \kappa >0$.

It is proven in Theorem 8.1 of \cite{21} that we have an eigenvalue  at $k=i\kappa$ for some positive $\kappa$ if
and only if $\text{Ker}[J(i\kappa)]$ is nontrivial
or equivalently if and only if $\text{det}[J(i\kappa)]=0.$ The multiplicity, $m_\kappa,$ of the eigenvalue at $k=i \kappa$ is finite and it is equal to the dimension of    $\text{Ker}[J(i\kappa)]$   Furthermore,  it is proven in Theorem 8.5 of \cite{21}  that,

\beq\label{3.19}
\det J(k)=c_3(k-i\kappa)^{m_\kappa}[1+O(k-i\kappa)],\qquad
k\to i\kappa,
\ene
where $c_3$ is a nonzero constant. Consequently, the order of the zero of $\det J(k)$ at $k=i\kappa$ is
equal to the multiplicity, $m_\kappa,$ of the eigenvalue at   $k=i\kappa$. We denote by $\mathcal N$ the total number of eigenvalues of $H_{A,B}$, repeated according to its multiplicity. If \eqref{1.6} holds, $ \mathcal N < \infty$.

\section{Trace Identities} \sss

We first obtain a high-energy asymptotic expansion for the Jost solution. Let us denote,

\beq \label{4.1}
m(k,x):= e^{-i k x}\, f(k,x).
\ene

\begin{lemma} \label{lemm4.1}
Suppose that the matrix potential $V$  is selfadjoint, that it belongs to $C^\infty((0,\infty))$ and that it satisfies \eqref{1.7}.
Then, for any $N=0,1,2,\cdots,$

\beq \label{4.3}
m(k,x)= \sum_{l=0}^N\, \frac{1}{(2 ik)^l}\, b_l(x)+ r_N(k,x), 
\ene
where the remainder  $r_N$ satisfies the estimate: for every $C >0$ there are constants $C_{N,j}$ such that,
\beq \label{4.4}
\left|  \frac{\partial ^j}{\partial x^j}\,  r_N(k, x) \right| \leq \, C_{N,j}\, \frac{1}{|k|^{N+1}}\, \frac{1}{(1+|x|)^{(N+1) (\rho-1)+j}}, \qquad j=0,1,2,\cdots, x \in [0,\infty),\, \hbox{\rm Im} k \geq 0, |k|\geq C,
\ene
where  $b_l(x), l=0,1,2,\cdots$ are $C^\infty$ functions defined by the following recurrent relation,
\beq \label{4.5}
 b_0(k,x)=1,\qquad    b_{l+1}(x)= - b'_l(x)- \int_x^\infty\, V(y)\, b_l(y)\, dy.
\ene
The $b_l(x)$ satisfy the estimate,
\beq \label{4.6}
\left|  \frac{d^j}{d x^j}\, b_l(x)  \right| \leq C_{l,j} \frac{1}{|x|^{l (\rho-1)+j}}, \qquad j=0,1,2,\cdots, l=0,1,2,\cdots.
\ene 
  Furthermore the expansion \eqref{4.3} can be derived term by term.
 \end{lemma} 
  \begin{remark}\label{rm4.2}
{ \rm  When an  asymptotic expansion as \eqref{4.3} holds for all $N$ we will write the right-hand side as an asymptotic series (see \cite{er})   }
\beq \label{4.7}
m(k,x)= \sum_{l=0}^ \infty\,  \frac{1}{(2 ik)^j}\, b_l(x).
\ene
\end{remark}

\noindent {\it Proof:} Results of this type are classical. For the reader's convenience we outline the proof following the one given in the scalar case  in Proposition 4.1 in Chapter 4 of \cite{ya2}, with some changes. The coefficients $b_l, l \geq 1$ are defined by the recurrent formula \eqref{4.5} and one easily checks that \eqref{4.6} is satisfied. Inserting \eqref{4.1} and \eqref{4.3} into \eqref{3.1} we obtain the equation for the remainder,
\beq \label{4.8}
- r''_N(k,x)-2 i \,k \,r'_N(k,x)+ V(x)\, r_N(k,x)= q_N(k,x),
\ene
where,
\beq \label{4.9}
q_N(k,x):= \left( b''_N(x)- V(x) b_N(x)  \right) \, \frac{1}{(2i k)^N}.
\ene
By  (\ref{1.7}, \ref{4.6})
\beq \label{4.10}
\left| \frac{\partial^j }{\partial x^j} q_N(k,x)  \right| \leq C_{N,j} \, \left( 1+x \right)^{-(N+1)(\rho-1)-1-j}\, \frac{1}{(2i k)^{N}}.
\ene
Writing \eqref{4.8} as an integral equation we get,
\beq \label{4.11}
r_N(k,x)= Q_N(k,x)+ \frac{1}{2 i k}\, \int_x^\infty\, \left(  e^{2 i k (y-x)} -1 \right)\, V(y)\, r_N(k,y) \, dy,
\ene
where,
\beq \label{4.12}
Q_N(k,x):= \frac{1}{2 i k}\,  \int_x^\infty\, \left( 1- e^{2 i k (y-x)} \right)\, q_N(k,y) \, dy.
\ene
By \eqref{4.10}
\beq \label{4.13}
\left| Q_N(k,x)  \right| \leq C_{N} \, \left( 1+x \right)^{-(N+1)(\rho-1)}\, \frac{1}{(2i k)^{N+1}}.
\ene 
Then, solving \eqref{4.11} by iterations and using \eqref{4.13} we prove that $r_N(k,x)$ satisfies \eqref{4.4} with $j=0$. To prove it for positive $j$ we proceed as follows (in the proof of Proposition 4.1 in Chapter 4 of \cite{ ya2} in  the scalar case a different argument is proposed). We assume that \eqref{4.4} is true for $j=0,1,\cdots j_0$ and we prove it for $j=j_0+1$. We denote $p(k,x):= \frac{\partial^{j_0+1}}{\partial x^{j_0+1}}\, r_N(k,x)$. Then, deriving \eqref{4.8} $j_0+1$ times we get that,
\beq \label{4.14}
- p''(k,x)-2 i \,k \,p'(k,x)+ V(x)\, p(k,x)= s(k,x),
\ene
where,
\beq \label{4.15}
s(k,x)= -\sum_{j=1}^{j=j_0+1}\,  \binom{j_0+1}{j}\, \left(  \frac{d x^j}{dx^j} V(x)\right)\, \frac{ \partial^{j_0+1-j}}{\partial x^{j_0+1-j}}\, r_N(k,x) +  \frac{ \partial^{j_0+1}}{\partial x^{j_0+1}}\, q_N(k,x).
\ene
By \eqref{1.7}, \eqref{4.10} and the inductive assumption we have that,

\beq\label{4.16}
\left| s(k,x)  \right| \leq \, C \left( 1+x \right)^{-(N+1)(\rho-1)-j_0-2}\, \frac{1}{(2i k)^{N}}.
\ene
Writing \eqref{4.14} as an integral equation we obtain that,
 \beq \label{4.17}
p(k,x)= S(k,x)+ \frac{1}{2 i k}\, \int_x^\infty\, \left(  e^{2 i k (y-x)} -1 \right)\, V(y)\, p(k,y) \, dy,
\ene
where,
\beq \label{4.18}
S(k,x):= \frac{1}{2 i k}\,  \int_x^\infty\, \left( 1- e^{2 i k (y-x)} \right)\, s(k,y) \, dy.
\ene
By \eqref{4.16} we have that $S(k,x)$ satisfies 

  \beq\label{4.19}
\left| S(k,x)  \right| \leq \, \left( 1+x \right)^{-(N+1)(\rho-1)-j_0-1}\, \frac{1}{(2i k)^{N+1}}.
\ene
Finally, solving   \eqref{4.17} by iterations and using \eqref{4.19} we obtain that,
  \beq\label{4.20}
\left| p(k,x)  \right| \leq \, \left( 1+x \right)^{-(N+1)(\rho-1)-j_0-1}\, \frac{1}{(2i k)^{N+1}},
\ene
 what proves that \eqref{4.4} holds for $r_N(k,x)$ with $j=j_0+1$.
  
  \bull
  
  By Lemma \ref{lemm4.1} the Jost matrix $J(k)$ defined in \eqref{3.7} (see also \eqref{3.8})  has the following asymptotic expansion,
  
  \beq \label{4.21}
  J(k)= f(-k^*,0)^\dagger\, B - f'(-k^*,0) ^\dagger\, A= \sum_{l=-1}^\infty \, c_l\, \frac{1}{(2ik)^l},
  \ene
  where,
  \beq\label{4.22}
  \begin{split}
  c_{-1} &= -\frac{1}{2}\, A ,\\
  c_0 &= B- \frac{1}{2}\,  b_1(0)^\dagger \,A, \\
  c_l &= b_l(0)^\dagger B -\left( \frac{1}{2} b_{l+1}(0)^\dagger ] +b'(0)^\dagger_{l}\right) A, l = 1,2,\cdots.
  \end{split}
  \ene
  Let us denote by $\mathcal P$ the set of all permutations of $1,2,\cdots, n$. By \eqref{4.21} we have that
  \beq\label{4.23}
  \det J(k)= \sum_{l=-n}^\infty  d_l\, \frac{1}{(2i k)^l},
  \ene 
  with
  \beq \label{4.24}
  d_l:= \sum_{\sigma \in \mathcal P} \,\mathrm{sign} \,\sigma  \sum_{l_i \geq -1: l_1+l_2+\cdots l_n=l} \, (c_{l_1})_{1,\sigma_1}\, (c_{l_2})_{2,.\sigma_2} \cdots    (c_{l_n})_{n, \sigma_n},
  \ene
  where by $(c_l)_{j,m}$ we denote the component of the matrix $c_l$ in the row $j$ and the column $m$.    
  
  By \eqref{3.18}, 
  \beq\label{4.25}
  d_j=0 \, \hbox{\rm if} \,  j < - n_M-n_N, \qquad \hbox{\rm and}\,  d_{-n_M-n_N}= \frac{c_2}{(2i)^{n_M+n_N}}.
  \ene
  Then, we have proven the following lemma.
  
  \begin{lemma}\label{lemm4.3}
 Suppose that the matrix potential $V$  is selfadjoint,  that it belongs to  $C^\infty((0, \infty))$ and that it satisfies \eqref{1.7}. Then, the  determinant of the Jost matrix, defined in \eqref{3.7}, has the following asymptotic expansion
 \beq\label{4.26}
  \det J(k)= \sum_{l=-n_M-n_N}^\infty  d_l\, \frac{1}{(2i k)^l},
  \ene 
where the coefficients $d_l, l=-n_M-n_N, -n_M-n_N+1,\cdots$ are defined in (\ref{4.5}, \ref{4.22}, \ref{4.24}, \ref{4.25} ) with $c_2$ the constant that appears in \eqref{3.18} and $n_M, n_N$  the nonnegative integers defined after equation \eqref{3.13}.
\end{lemma}
 Let us denote,
 \beq\label{4.27}
 h(k):= \frac{1}{c_2\, k^{n_M+n_N}}\, \det J(k).
 \ene
  By \eqref{3.18},
  $\lim_{|k|\rightarrow \infty} \, h(k)=1.$
  Hence, we define $\ln h(k)$ with $\lim_{|k| \rightarrow \infty} \, \ln h(k)=0$. By \eqref{4.26}
  \beq \label{4.28}
 \ln  h(k) = \sum_{l=1}^\infty\, e_l\, \frac{1}{(2 i k)^l},
 \ene
 where,
 \beq \label{4.29}
 \begin{split}
  e_{1} &= \frac{(2 i)^{n_M+n_N}}{c_2}\, \, d_{1-n_M-n_N} ,\\
  e_l &=     \frac{(2 i)^{n_M+n_N}}{c_2}\, \left( d_{l-n_M-n_N} - \frac{1}{l}\, \sum_{j=1}^{l-1}\,  j   \,d_{l-n_M-n_N-j} \,\,e_{j}\, \right), l= 2,\cdots. 
  \end{split}
  \ene
   For $k$ real we split  the asymptotic expansion \eqref{4.28} into its even and odd parts,
\beq\label{4.30}
\ln h(k)= \ln_e h(k)+ \ln_o (k),\, \ln_e h(k):= \frac{1}{2} \left( \ln h(k)+\ln  h(-k)\right), \quad \ln_o h(k):=  \frac{1}{2} \left(\ln h(k)-\ln h(-k)\right), \quad k \in \mathbf{R}.
\ene
By \eqref{4.28},
\beq \label{4.31}
  \ln_e h(k)   = \sum_{l=1}^\infty\, e_{2l}\, \frac{1}{(2 i k)^{2l}}=   \sum_{l=1}^\infty\, (-1)^{l}  e_{2l}\, \frac{1}{(2  k)^{2l}}, \quad k \in \mathbf{R},
  \ene
  \beq \label{4.32}
  \ln_o h(k):=  \sum_{l=0}^\infty\, e_{2l+1}\, \frac{1}{(2 i k)^{2l+1}}= i \sum_{l=0}^\infty\,  (-1)^{l+1}\, e_{2l+1}\, \frac{1}{(2  k)^{2l+1}},  \quad k \in \mathbf{R}.
  \ene 

\begin{lemma}\label{lemm3.4}
Suppose that the selfadjoint matrix potential $V$  satisfies  \eqref{1.6}. For any $ z \in \mathbf C$ with $ 0 < {\rm Re} z < 1/2$ let us define,
\beq \label{4.33}
F(z):= \int_0^\infty\, \ln_e h(k)\, k^{2z-1}\, dz, \qquad G(z):=- i\, \int_0^\infty \, \ln_o h(k)\, k^{2z-1}\, dz.
\ene
Then,
\beq \label{4.34}
 F(z)\, \sin(\pi z) - G(z)\, \cos(\pi z)=     \frac{\pi}{2 z} \, \widetilde{\sum}_{j=1}^\mathcal N \left|  \kappa_j \right|^{2z},
 \ene
 where $- \kappa_j^2, j=1,2\cdots, \mathcal N$ are the eigenvalues  of the Schr\"odinger operator $H_{A,B}$ defined in \eqref{1.1} with the boundary conditions (\ref{1.2}- \ref{1.4}) . In the right-hand side of \eqref{4.34}    $\widetilde{\sum}_{j=1}^\mathcal N    \left|  \kappa_j \right|^{2z}$  means the sum over the absolute value of the eigenvalues,
 $-  \kappa_j^2$, to the power $z$ and    repeated according to its multiplicity. Furthermore,  $\mathcal N$ is the total number of  (repeated)  eigenvalues. As \eqref{1.6} holds, $ \mathcal N < \infty.$
 \end{lemma}

\noindent{\it Proof:} This lemma is proven as in the proof of Proposition 6.3 in Chapter 4 of \cite{ya2} integrating the function \linebreak
$ \left(\dot{h}(k)/ h(k)\right) \, k^{2z}$, with $h(k)$ defined in \eqref{4.27} with 
$0 \leq \textrm{arg} k \leq \pi$ along the contour $\mathcal C_{\epsilon,R},$ given below,  taking the limit when $ \epsilon \rightarrow 0$ and $  R \rightarrow \infty$ and using 
\eqref{3.17}, \eqref{3.18},  \eqref{3.19} and the residues theorem.  The contour  $\mathcal C_{\epsilon,R}$ consists of four parts   given by

$$\mathcal C_{\epsilon,R}:=(-R,-\epsilon)\cup\mathcal C_\epsilon\cup(\epsilon,R)\cup\mathcal C_R.
$$

The  piece $(-R,-\epsilon)$ is the directed line segment on the real axis
for some small positive $\epsilon$ and for a large positive $R,$
with the direction of the path from $-R+i0$ to $-\epsilon+i0.$
The second part $\mathcal C_\epsilon$ consists of the
upper semicircle centered at the origin with radius
$\epsilon$ and traversed from the point $-\epsilon+i0$ to
the point $\epsilon+i0.$ The third piece $(\epsilon,R)$
is the directed line segment of the positive real axis from $\epsilon+i0$ to $R+i0.$
The fourth part $\mathcal C_R$ is the
upper semicircle centered at the origin with radius
$R$ and traversed from the point $R+i0$ to
the point $-R+i0.$

\begin{prop}\label{prop3.5}
Assume that the selfadjoint matrix potential  $V$ satisfies \eqref{1.6}, that it   belongs to $C^\infty((0,\infty))$ and that \eqref{1.7} holds. Then, the functions $F(z)$ and $G(z)$, defined in \eqref{4.33} have  analytic continuations, denoted also by $F(z), G(z)$, to  meromorphic functions for $ \hbox{\rm Re} z > 0.$ The function $F(z)$ has simple poles at $z=j, j=1,2,\cdots$ with residue $(-1)^{j+1}\,  2^{-2j-1} \, e_{2j}$. Furthermore, the representation \eqref{4.33} of $F(z)$ is valid for $ 0 <\hbox{\rm Re} \, z < 1,$ and,
\beq \label{4.35}
F(z)= \int_0^\infty \left( \ln_e h(k)\, -\sum_{l=1}^j (-1)^l \, e_{2l}\, \frac{1}{(2k)^{2l}} \right)\, k^{2z-1} \, dk, \quad \hbox{\rm for} \, j < \hbox{\rm Re}\, z < j+1, j=1,2,\cdots.
\ene
Moreover, $G(z)$ has simple poles at $z=j+1/2, j=0,1,2,\cdots$ with residue, $ (-1)^j\,  2^{-2 j-2}\, e_{2 j+1}$ and  
\beq \label{4.36}
G(z)= \int_0^\infty\,  \left( -i \ln_o h (k)- \sum_{l=0}^{j-1}\,  (-1)^{l+1}\, e_{2l+1}\, \frac{1}{(2k)^{2l+1}} \right)\, k^{2z-1}\, dk, \qquad j-\frac{1}{2}  < \hbox{\rm Re}\, z < j+\frac{1}{2}, j=1, 2,\cdots.
\ene
\end{prop}
\noindent {\it Proof:}      The proposition follows from \eqref{4.31}, \eqref{4.32} as in the proof of Lemma 6.4 in Chapter 4 of \cite{ya2}. We give details in the case of $G(z)$ for the reader's convenience. We have that for $ \hbox{Rez } < \frac{1}{2},$
\begin{eqnarray} \label{4.37}  \nonumber
G(z)= \int_0^1 (-i) \ln_o h(k)\,  k^{2z-1}\, dk +\int_1^\infty \left( -i \ln_o h(k)- \sum_{l=0}^{j-1} \, \frac{(-1)^{l+1} e_{2l+1}}{(2k)^{2l+1}} \right )\, k^{2z-1}\, dk \\
 +  
 \sum_{l=0}^{j-1} \, (-1)^{l} \, 2^{-2l-2} \frac{e_{2 l+1}} {z-l- 1/2}.
 \end{eqnarray}
The first integral in the right-hand side of \eqref{4.37} is analytic for $ \hbox{Re} z >0$ and by \eqref{4.32} the second  integral is analytic for $ \hbox{\rm Re } z < j+1/2$. Since this is true for every $j=1,2,\cdots,$ $G(z)$ has an analytic continuation to a meromorphic function for $ \hbox{Re}\, z >0$ with simple poles at   $z=j+1/2, j=0,1,2,\cdots,$ with residue, $ (-1)^j\,  2^{-2 j-2}\, e_{2 j+1}$. Moreover, for  $j-\frac{1}{2}  <    \hbox{Re} z,$
\begin{eqnarray} \label{4.38} \nonumber
  \int_0^1 (-i) \ln_o h(k)\, k^{2z-1}\, dk = \int_0^1 \left( -i \ln_o h(k)- \sum_{l=0}^{j-1} \, \frac{(-1)^{l+1} \,  e_{2l+1}}{(2k)^{2l+1}} \right )\, k^{2z-1}\, dk  \\  -
   \sum_{l=0}^{j-1} \, (-1)^{l} \, 2^{-2l-2} e_{2l+1} \, \frac{1}{z-l- 1/2}.     
    \end{eqnarray}
   Equation \eqref{4.36} follows from \eqref{4.37} and \eqref{4.38}.
   
   \bull

We now prove our main result.
\begin{theorem}\label{theor3.6}
Suppose that the selfadjoint matrix potential $V$ satisfies \eqref{1.6} that it belongs to $C^\infty ((0,\infty))$  and that  \eqref{1.7} holds. Then,
\beq \label{4.39}
  \widetilde{\sum}_{l=1}^\mathcal N \left|  \kappa_l \right| - \frac{1}{\pi} \int_0^\infty\,  \ln_e h(k) \, dk = \frac{e_1}{4},
  \ene
  \beq \label{4.40}
   \widetilde{\sum}_{l=1}^\mathcal N \left|  \kappa_l \right|^{2j+1}+ (-1)^{j+1} \frac{2j+1}{\pi}\, \int_0^\infty \left( \ln_e h(k)\, -\sum_{l=1}^j (-1)^l \, e_{2l}\, \frac{1}{(2k)^{2l}} \right)\, k^{2j} \, dk = \frac{(2j+1) \,e_{2j+1} }{2^{2j+2}}, j=1,2,\cdots.
   \ene
 Furthermore, 
  \beq \label{4.41}
   \widetilde{\sum}_{l=1}^\mathcal N \left|  \kappa_l \right|^{2j}+ (-1)^{j} \frac{2j}{\pi}\, \int_0^\infty \left(-i \ln_o h(k)\, -\sum_{l=0}^{j-1} (-1)^{l+1} \, e_{2l+1}\, \frac{1}{(2k)^{2l+1}} \right)\, k^{2j-1} \, dk =-j \frac{ e_{2j} }{2^{2j}}, j=1,2,\cdots.
 \ene  
  In the left -hand side of \eqref{4.39}, \eqref{4.40} and \eqref{4.41} $\widetilde{\sum}_{l=1}^\mathcal N    \left|  \kappa_l \right|^{q}$,  with, respectively, $q=1,q=2j+1,$ and
   $q=2j,$   means the sum over the absolute value of the eigenvalues,
 $-  \kappa_l^2$, to the power $q/2$, and    repeated according to its multiplicity. Furthermore,  $\mathcal N$ is the total number of  (repeated)  eigenvalues. As \eqref{1.6} holds, $\mathcal N < \infty.$ The coeficients $e_j, j=1,2,\cdots$ are defined in equations   (\ref{4.5}, \ref{4.22}, \ref{4.24}, \ref{4.25}, \ref{4.29}).
 \end{theorem}
 
   \noindent{\it Proof:} By Proposition \ref{prop3.5} and analytic continuation \eqref{4.34} holds for $ \hbox{\rm Re} z > 0$. Equations \eqref{4.39} and \eqref{4.40} follow evaluating
   \eqref{4.34} at $z=j+1/2, j=0,1,2,\cdots$ and using \eqref{4.35}. Moreover, \eqref{4.41} follows evaluating \eqref{4.34} at $z= j, j=1,2,\cdots$ and using \eqref{4.36}.
   
   \section{Examples} \sss 
   In this section we illustrate our trace identities in Theorem \ref{theor3.6} with simple examples.
   \begin{example} \label{ex5.1}{ \rm
   We consider a $ 2 \times 2$ system, i.e. $n=2,$ with Dirichlet boundary condition, $\psi(0)=0$. We can take, $A=0, B=-I$.  Since we only have Dirichlet  boundary conditions according to the definition given below equation \eqref{3.13}  $n_M=n_N=0$ and also the constant $c_2$ that appears in \eqref{3.18} is equal to one (see the first equation in page 16 of \cite{21}). Then according to   (\ref{4.5}, \ref{4.22}, \ref{4.24}, \ref{4.25}, \ref{4.29}).
   
   \beq \label{5.1}
   e_1= -\int_0^\infty
\, V_{1,1}(x)\, dx -    \int_0^\infty
\, V_{2,2}(x)\, dx.
\ene
\begin{eqnarray}\nonumber
e_2&=&\left(  \int_0^\infty \, V_{1,1}(x)\, dx \right)\, \left(  \int_0^\infty \, V_{2,2}(x)\, dx \right) - \left(  \int_0^\infty \, V_{1,2}(x)\, dx \right)\, \left(  \int_0^\infty \, V_{2,1}(x)\, dx \right)
- V_{1,1}(0)
\\     \nonumber
& +& \left(\int_0^\infty\, dx \, \int_x^\infty\, V(y)\, dy \, V(x)\right)_{1,1}
- V_{2,2}(0) + \left(\int_0^\infty\, dx \,  \int_x^\infty\, V(y)\, dy \, V(x)\right)_{2,2} 
\\ \label{5.2}
&-& \frac{1}{2} \left(   \int_0^\infty
\, V_{1,1}(x)\, dx +    \int_0^\infty
\, V_{2,2}(x)\, dx \right)^2.
\end{eqnarray}
   }  \end{example}

\begin{example}{(The $\delta'$ boundary condition) }. \label{ex5.2} {\rm
We consider a  $3 \times 3$  system i.e., $n=3$. It satisfies the $\delta'$ boundary condition,
\beq \label{5.3}
\psi'_1(0)= \psi'_2(0)= \psi'_3(0), \qquad \sum_{j=1}^3  \psi_j(0)  = a \psi'_1(0), a \in \mathbf R.
\ene
The matrices $A,B$ can be taken as, 
 $$
 A=\bm 1&0&-a\\
\stretch
-1&1&0\\
\stretch
0&-1&0\endbm,\quad
B=\bm 0&0&-1\\
\stretch
0&0&-1\\
\stretch
0&0&-1\endbm.
$$
The potential $V$ is identically zero. The matrices $A,B$ satisfy \eqref{1.3}, \eqref{1.4}. From \eqref{3.8} with
$f(k,x)=e^{ikx}I_3 ,$ we obtain that,
$$
J_{A,B}(k)=
\bm -ik&0&-1+iak\\
\stretch
ik&-ik&-1\\
\stretch
0&ik&-1\endbm.$$
Then,
\beq \label{5.4}
\det J_{A,B}(k)= k^2 (3-ia k).
\ene
It follows that if $a \geq 0$ the Schr\"odinger operator $H_{A,B}$ has no eigenvalues. On the contrary, if $ a <0$,  $H_{A,B}$ has one eigenvalue $- \kappa_1^2$ with
$i \kappa_1= i  3 /|a|$ and with multiplicity one. 
}
{\rm
Suppose that $ a \neq 0$. Then,  with $h(k)$ defined in \eqref{4.27}
\beq \label{5.5}
h(k)= \frac{\det J_{A,B}(k)}{-i  a k^3}= \left( 1 +\frac{3i}{a k}\right).
\ene 
Then,
\beq \label{5.6}
\ln h(k)= \sum_{l=1}^\infty \frac{-1}{l} \, \left(\frac{6}{a}\right)^l \, \frac{1}{(2ik)^l}.
\ene
It follows that
\beq\label{5.7}
e_l=  \frac{-1}{l} \, \left(\frac{6}{a}\right)^l, l=1,2,\cdots .
\ene 
In the case $a=0$,  $h(k)=1$ and then $\ln h(k)=0$ and $e_l=0, l=1,2,3,\cdots.$
}
\end{example}

\end{document}